\documentclass[final,conference,10pt,a4paper]{IEEEtran}

\usepackage{url}
\usepackage[bookmarks, hidelinks]{hyperref} 
\usepackage{listings}
\usepackage{xspace}
\usepackage{graphicx}
\usepackage{url}
\usepackage[flushleft]{threeparttable}
\usepackage{amsfonts,amsmath,amscd,amssymb,array}
\usepackage{float}

\lstnewenvironment{mylisting}[1][]
	{\lstset{
		float=h,
		basicstyle=\ttfamily\scriptsize,
		captionpos=b,
		#1}
	}
{}

\newcommand{\nameit}{PALPAS\xspace}
\newcommand{\sss}{SSS\xspace}

\newcommand{\seed}{\ensuremath{S}\xspace}
\newcommand{\kmpw}{\ensuremath{K_{\mathrm{MPW}}}\xspace}
\newcommand{\ke}{\ensuremath{K_{\mathrm{E}}}\xspace}
\newcommand{\kauth}{\ensuremath{K_{\mathrm{Auth}}}\xspace}
\newcommand{\kauthtwo}{\ensuremath{K'_{\mathrm{Auth}}}\xspace}
\newcommand{\tauth}{\ensuremath{T_{\mathrm{Auth}}}\xspace}

\newcommand{\no}{\centering \ensuremath{\circ}}
\newcommand{\yes}{\centering \ensuremath{\bullet}}

\IEEEoverridecommandlockouts
\newcommand{\scale}{1}
	
\begin{document}

\title{PALPAS\\PAsswordLess PAssword Synchronization}

\IEEEspecialpapernotice{Extended Version}

\author{
	\IEEEauthorblockN{Moritz Horsch}
	\IEEEauthorblockA{
		Technische Universit{\"a}t Darmstadt\\
		Hochschulstra{\ss}e 10\\64283 Darmstadt, Germany\\
		horsch@cdc.informatik.tu-darmstadt.de
	}
	\and
	\IEEEauthorblockN{Andreas H{\"u}lsing}
	\IEEEauthorblockA{
		Technische Universiteit Eindhoven\\
		5600 MB Eindhoven, Netherlands\\
		{a.t.huelsing@tue.nl}
	}
	\and
	\IEEEauthorblockN{Johannes Buchmann}
	\IEEEauthorblockA{
		Technische Universit{\"a}t Darmstadt\\
		Hochschulstra{\ss}e 10\\64283 Darmstadt, Germany\\
		buchmann@cdc.informatik.tu-darmstadt.de
	}
	\thanks{An extended abstract of this work appears in the proceedings of ARES 2015.}
}

\maketitle   

\thispagestyle{plain}
\pagestyle{plain}

\begin{abstract}
Tools that synchronize passwords over several user devices typically store the encrypted passwords in a central online database. For encryption, a low-entropy, password-based key is used. Such a database may be subject to unauthorized access which can lead to the disclosure of all passwords by an offline brute-force attack. In this paper, we present PALPAS, a secure and user-friendly tool that synchronizes passwords between user devices without storing information about them centrally.  The idea of PALPAS is to generate a password from a high entropy secret shared by all devices and a random salt value for each service. Only the salt values are stored on a server but not the secret. The salt enables the user devices to generate the same password but is statistically independent of the password. In order for PALPAS to generate passwords according to different password policies, we also present a mechanism that automatically retrieves and processes the password requirements of services. PALPAS users need to only memorize a single password and the setup of PALPAS on a further device demands only a one-time transfer of few static data.
\end{abstract}

\section{Introduction}
The concept of using passwords for user authentication on the Internet is cost-effective for services and easily comprehensible for users. However, the key challenge for users is to choose a strong password for each service and never reuse it for another service. This is essential for the security, but the wide adoption of passwords on the Internet makes it impossible for users to memorize the required amount of different strong passwords. To mitigate this conflict of having user-friendly and secure passwords one possible solution is storing some data locally on the user's device, which can be the passwords themselves or information to compute them.

However, this requires an additional protection mechanism to ensure that only the legitimate user is able to access the data, e.g., by a proof of knowledge or possession. Another problem of storing data locally is that it is only available on a single device but nowadays users have multiple devices like smart phones and tablets. Furthermore, the data is not static. Passwords get added, changed, or removed. Thus, the data must be synchronized between devices to ensure that users can access their services at any time and any place. The data can be synchronized manually by directly connecting the devices or in a more convenient and user-friendly way through a synchronization server over the Internet.

There exist many solutions for the generation of individual passwords, secure password storage, and password synchronization, but they have serious drawbacks. The common paradigm of password synchronization is protecting passwords via encryption with a password-derived key and storing them at servers on the Internet. This bears the risk of security breaches at the synchronization servers where adversaries copy the encrypted data and perform offline attacks (cf. \cite{DBLP:conf/esorics/GastiR12,lastpasssn}). Hence, there is a need for a synchronization scheme that is on the one hand not vulnerable to such attacks and on the other hand user-friendly with respect to the amount of information users need to memorize and the number of actions users need to perform to synchronize their passwords.

The protection of local and/or remote data as well as the user authentication at the synchronization server are usually based on a user-chosen (master) password. This enables various attacks based on web vulnerabilities and phishing attacks in which users are tricked into submitting their password to an adversary (cf. \cite{DBLP:conf/woot/BhargavanD12,DBLP:conf/ccs/StockJ14}). Hence, there is a need for a solution that strictly separates the duties of providing data and privacy protection as well as user authentication.

Existing approaches to generate random passwords do not consider the password requirements of services, but this is crucial to create strong passwords. Usually, a default character set and password length are used, which hopefully fit the requirements of all services. Otherwise, users need to adapt the password or configure the password generator for the particular service. In both cases, users need to find the password requirements at the service's website or end up in an exhausting trial and error approach to find out which kinds of passwords are accepted. This situation is very inconvenient and leads users to use the first password accepted by the service and not the strongest and most appropriate one. Hence, there is a need for an automatic mechanism that allows password generators to retrieve and process the password requirements of services without requiring any user interaction.

In this work we present PALPAS, a novel password tool that creates strong, service-specific passwords and synchronizes them between devices via a central server. However, it does not store or use any passwords on the server and is therefore not vulnerable to phishing attacks or security breaches.

More specifically, we make the following contribution regarding the open issues described above:

\begin{itemize}
	\item We present a synchronization scheme that shares data between devices through a synchronization server to enable them to compute the same passwords. However, the data is statistically independent of the passwords and hence does not reveal any information about them. Moreover, the user authentication at the synchronization server is not based on passwords but on public key cryptography. Therefore, a security breach at the synchronization server is non-critical, because there are no passwords that could be stolen by an adversary.
	
	\item We describe an approach to use different, independent high entropy secrets for password generation, data and privacy protection, and user authentication, which provides a maximum of security on the one hand but is still easily manageable and user-friendly on the other hand.
	
	\item We provide a common description of password policies to specify the password requirements of services in a standardized way. We describe a solution to create and distribute the policies. This allows us to finally provide a mechanism which automatically retrieves the password requirements of arbitrary services to create strong and service-specific passwords. It requires no user interaction, except that users need to specify for which service the password should be created.
\end{itemize}

This paper is organized as follows: In Section \ref{sec:relatedWork} we summarize related work. We describe \nameit on a conceptual level in Section \ref{sec:PALPAS} and provide more details about its implementation in Section \ref{sec:implementation}. In Section \ref{sec:security} we present a detailed security analysis of our concept and finally conclude the paper in Section \ref{sec:conclusion}. The appendix includes detailed protocol flows.

\section{Related Work}
\label{sec:relatedWork}

Password managers like KeePass \cite{keepass} and LastPass \cite{lastpass} as well as the browser built-in password managers of Chrome and Firefox \cite{fxa} encrypt the passwords before storing them at servers on the Internet. The encryption key is derived from a user-chosen master password. A security breach at a synchronization server (cf. \cite{lastpasssn}) would allow adversaries to steal the encrypted data and to perform offline brute-force attacks. In essence, protecting passwords by another password and storing them on servers cannot be the foundation for a secure password synchronization.   

Hash-based approaches like PwdHash \cite{DBLP:conf/uss/RossJMBM05} allow users to create different passwords for services by hashing a master password and the name or the URL of the service. Unfortunately, an adversary who stole a password can perform a brute-force attack to obtain the master password and thereby generate all user's passwords. Password Multiplier \cite{DBLP:conf/www/HaldermanWF05} performs additional steps to strengthen the master password, which increases the costs for a brute-force attack. Nevertheless, these hash-based approaches cannot be used to generate a new password for the same service, which is e.g.\ necessary after a password breach. The authors propose to use an additional user-chosen input for each service, which is included in the hashing, but this has the disadvantage that users then need to memorize this input for each service. Thus, the existing hash-based approaches are not a feasible solution because they still require users to memorize a lot of information.

Approaches which are using hardware tokens \cite{DBLP:conf/spw/StajanoJPSSW14,DBLP:conf/nss/VarmedalKHJVM13} or mobile devices \cite{DBLP:journals/tifs/SunCL12} for authentication have the disadvantage that users always need to carry an additional device. Furthermore, such solutions require changes on the infrastructure of the service. The weak development of other authentication mechanisms than passwords (cf. \cite{DBLP:conf/sp/BonneauHOS12} for a survey) shows that service-side changes are a major obstacle for the wide adoption of authentication schemes.

Single sign-on (SSO) like Facebook Connect \cite{fbc} allows users to authenticate themselves with a single password once and access multiple services without being prompted to log in at each service again. This can reduce the number of passwords users have to memorize but the adoption of SSO is still very limited \cite{DBLP:conf/nspw/SunBHB10}. Furthermore, SSO bears the risk of phishing attacks \cite{DBLP:conf/leet/Yue13} and studies found out that users have several concerns and misconceptions about SSO and are not feeling comfortable with giving control of their passwords to external services \cite{DBLP:conf/soups/SunPMDHB11,DBLP:journals/toit/SunPMDHB13}. SSO has also serious privacy issues \cite{DBLP:conf/chi/Egelman13}, because the SSO identity provider is aware of where and when a user performs a login. In summary, SSO does not solve the problem of managing many passwords and leads to new problems like the privacy issue.

\section{PAsswordLess PAssword Synchronization}
\label{sec:PALPAS}

In this section we describe \nameit on a conceptual level. We explain how \nameit creates passwords for services, how the individual password requirements of each service are met during the password generation, how the passwords are synchronized between several user-devices, and finally how the various data is protected from unauthorized access. Technical details about the implementation, the used encryption schemes, the used key sizes, and so forth are presented in Section \ref{sec:implementation}. Detailed protocol flows can be found in the Appendix.

\begin{figure}[b]
	\centering
	\includegraphics[width=.9\linewidth]{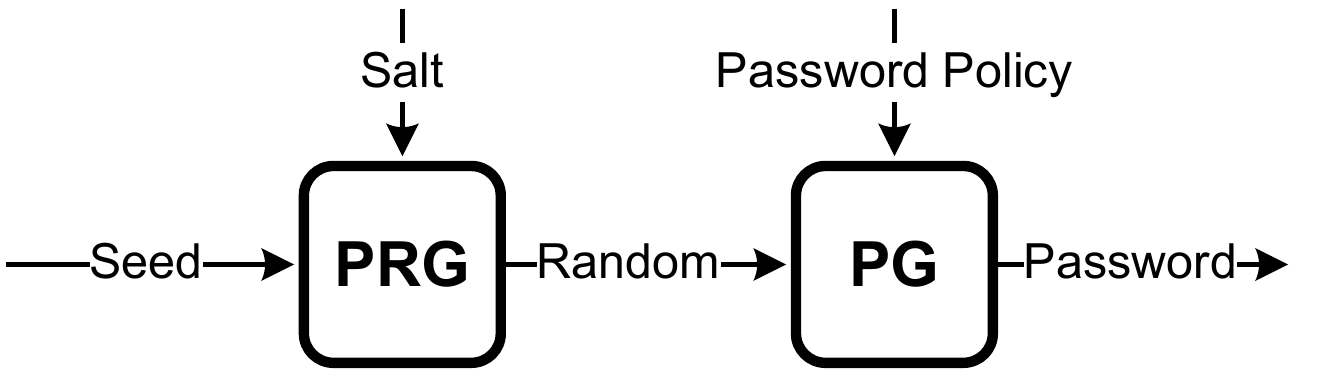}
	\caption{Password Generation. The Pseudorandom Generator (PRG) generates randomness based on a seed and a salt value. The Password Generator (PG) derives a password from the randomness and ensures that it complies with a password policy specifying the password requirements of the service.}
	\label{fig:Password_Generatior}
\end{figure}

\subsection{Password Generation}
\label{sec:ConceptPasswordGeneration}
\nameit does not store a password as typical password managers do. It (re-)computes a password every time the user wants to log in to a service. \nameit uses completely random passwords but ensures that they comply with the password requirements of a service. It also uses a different password for each service and never reuses it. All that together allows to have an individual and strong password for each service. 

As illustrated in Figure \ref{fig:Password_Generatior}, a password is computed in two steps: First, a (cryptographically secure) Pseudorandom Generator (PRG) generates a pseudorandom value based on a seed and a salt value. Second, a Password Generator (PG) derives a password from the pseudorandom value and ensures that it complies with a given password policy. The password policy specifies the password requirements of the service. The PRG and the PG are deterministic, thus, using the same seed, salt, and password policy, the same password is generated. We present more details about the implementation of the PRG and the PG in Section \ref{sec:ImplPseudorandomGenerator} and \ref{sec:ImplPasswordGeneration}.

The seed \seed is a common secret shared by all user devices. It is randomly generated when the user uses \nameit the first time and it must be transferred by the user to all his or her devices, but only once. In Section \ref{sec:Secrets} we provide more details how to do this in a convenient way. To protect the seed from unauthorized access it is encrypted using a secret \kmpw derived from a user-chosen master password (MPW). It is important to stress that this secret is only used for the local data protection and local user authentication. More precisely, it ensures that only the user is able to access the seed. The secret \kmpw is neither stored on a server nor used for any kind of data protection in which the data leaves the device. 

The objective of the salt is twofold. First, \nameit uses a different salt value for each service to create different passwords for the services. Second, changing the salt for a service allows generating a new password for it, which is necessary in case of a password breach or regular required password changes. The salt is chosen by \nameit and completely random. Consequently, it is statistically independent of the password and its knowledge does not allow to infer any user-, service-, or password-related information.

\subsection{Password Policies}
\label{sec:ConceptPasswordPolicies}
A password policy defines the password requirements of a service. It specifies the minimum and maximum password length, the allowed character sets, and additional constraints of the required characters. For instance, a password policy specifies that passwords must have at least 8 and a maximum of 10 characters, can consist of upper- and lowercase letters and digits, and must contain at least one digit.

\nameit uses a separate password policy for each service, but instead of putting the burden of creating these policies on the user, we provide an automatic mechanism to process and retrieve the password policies of arbitrary services. We propose a community approach to create the password polices and distribute them through a central service. This has the benefit that it is independent from the services and does not require their contribution. The basic concept is that users create a password policy for a service once and make it available for other users by distributing it through the central service.

We developed a common description for password policies to specify the password requirements of services in a standardized way (cf. Section \ref{sec:ImplPasswordPolicies} for an example). This builds the foundation to automatically retrieve and process password policies without any user interaction. We introduce the Password Policy Service (PPS) where applications like \nameit can automatically request the policy of an arbitrary service. The PPS provides a form for users where they can enter the password requirements of a service. The PPS then automatically creates the corresponding password policy and publishes it. To ensure the correctness of the password policy the PPS performs multiple verifications. It performs plausibility and sanity checks to ensure, for instance, that the generated passwords have at least a certain threshold of entropy. Moreover, it requires that at least several users enter the same password requirements of a service before a password policy gets published. In addition, the PPS provides a reputation system to allow users to rate the policies and to assist users in deciding whether to trust or not to trust in a certain policy. Applications like \nameit should also perform sanity checks to ensure that a generated password has a minimum level of entropy.

There is an issue when a service changes its password requirements but that only affects users that change the passwords. It requires that users submit the new password requirements to the PPS to update the password policy. \nameit retrieves the password policies for the services only once and stores them on the device. In case the user wants to have a new password for a service \nameit checks if there is a newer password policy to ensure that the password complies with the current password requirements. 

In summary, our approach enables applications to automatically retrieve and process password policies, without requiring users to manually search for the password requirements at the service's website and configure the application. The first \nameit users of a service still have to manually enter the requirements but we expect that after an initialization phase policies for all popular services are available. In Section \ref{sec:ImplPasswordPolicies} we provide more details regarding the implementation of password polices.

\begin{table*}[t]
	\renewcommand{\arraystretch}{1.2}
	\normalsize
	\small
	\centering
	\caption{Secrets used by \nameit}
	\label{table_example}
	\begin{tabular}{|p{.06\linewidth}|p{.2\linewidth}|p{.27\linewidth}|p{.35\linewidth}|}\hline
		\bfseries Secret & \bfseries Objective & \bfseries Generation & \bfseries Properties \\\hline\hline
		\seed & Password generation & - Random \newline - First use of \nameit & - Same for all devices\newline - Manually transfered by the user only once
		\\ \hline
		\kmpw & Local data protection and \newline local user authentication & - Derived from master password \newline - Every use of \nameit & - Same  
		for all devices\newline - Entered on each device \newline - Entered for every use of \nameit
		\\ \hline
		\ke & Remote data and privacy protection & - Random \newline - First use of \nameit &  - Same for all devices\newline - Manually transfered by the user only once
		\\ \hline
		\kauth & Device authentication &  - Random \newline - First use of \nameit on a device & - Different for all devices \newline 
		- Generated by each device\newline 
		- \tauth manually transfered only once
		\\\hline
	\end{tabular}
\end{table*}

\subsection{Password Synchronization}
\label{sec:ConceptPasswordSynchronization}
As described in Section \ref{sec:ConceptPasswordGeneration} a password for a service is computed from the seed, a salt, and a password policy. The seed is available on all user devices. The password policy can be retrieved from the PPS and is stored on the device after the first download. To enable all user devices to compute the same password for a service only the salt needs to be shared between the devices.

We introduce the Salt Synchronization Service (SSS), a central service that synchronizes the salt values between a user's devices. For each service or rather for each password of the user, the SSS stores a separate salt and an associated identifier. The identifier allows \nameit to request the salt for a particular service. In the interest of simplification, we assume that the identifier is the URL of the service. In case the user wants to log in to a service, \nameit requests the corresponding salt from the \sss and computes the password. A detailed protocol flow can be found in the Appendix.

This approach also allows \nameit to determine if a user already has an account at a particular service by checking if there is a salt for the service stored at the \sss. To synchronize a new password, \nameit just adds the new salt and the identifier to the \sss. After synchronizing with the SSS, all user devices are able to compute the password and to perform the login. The same applies if the user updates the password for a service, \nameit stores the new salt value using the old identifier at the \sss. This automatically makes the new password available to all other user devices.

Using the URL of a service as identifier leads to privacy issues as the \sss and potential adversaries would gain information about the services used by the user. Therefore, \nameit obtains the identifier hashing the URL with a secret value \ke. This secret key is randomly generated during the first use of \nameit. It must also be transferred by the user to all his or her devices when adding a new device. This can be done together with the seed.

Besides the password, a login to a service also demands a username. \nameit also provide means to synchronize the usernames between the user's devices. Usernames are sensitive information as they might reveal a user's real name on the one hand and give adversaries an advantage in social engineering attacks. On the other hand, knowledge of a username might allow to connect a user with his actions e.g. in forums or on dating sites. Thus, \nameit protects usernames by encrypting them using \ke from above and stores them on the \sss. 

In summary, for each service, used by the user, the \sss stores a triple consisting of the salt, the salt identifier, and the encrypted username.

\subsection{User Authentication at the SSS}
\label{sec:ConceptUserAuthentication}
We need to ensure that only the legitimate user is able to store, update, or delete his or her salt values at the \sss. Common approaches for user authentication at synchronization servers are password-based, which makes phishing attacks very likely. \nameit uses public key based authentication with an independent secret \kauth (and a corresponding public key) for the user authentication at the \sss. An authentication key pair is automatically generated on each user device.  When the user uses \nameit the first time, \nameit automatically generates a key pair and uses it to create an account at the \sss. This happens completely transparent to the user. The \sss only stores public keys and hence an attack against the \sss leaks no information usable to hijack a user account. If the user wants to set up \nameit on a further device, he or she requests an authentication token \tauth from the \sss, which is used by the new device to register its new authentication secret \kauthtwo at the \sss. This allows to revoke single authentication key pairs in case of loss. The authentication token \tauth is transfered together with seed \seed and secret \ke during set-up.

\subsection{Separation of Duties for Data and Privacy Protection and User Authentication}
\label{sec:Secrets}
\nameit needs to ensure that only the legitimate user is able to compute his or her passwords and to access the necessary data to do so. Furthermore, \nameit must protect any personal information about the user to ensure his or her privacy. As summarized in Table \ref{table_example}, \nameit uses four different secrets to achieve the goals. The seed ensures that only the legitimate user is able to compute the passwords, even if adversaries get access to the salt values. The secret \kmpw protects the data stored on devices and ensures that only the user who knows the MPW can access the local data. Usernames and salt identifiers might reveal personal information about the user; therefore they are protected by \ke before storing them on the \sss. Finally, to ensure that only the legitimate user can access (and hence edit) the data stored on the \sss, each device authenticates itself with a different secret \kauth. The seed \seed, the secret \ke, and the secret \kauth are randomly generated with sufficiently high entropy to make brute-force attacks meaningless. Only the secret \kmpw is derived from a user-chosen MPW. However, \kmpw is only used for local data protection and local user authentication and we use common techniques to make brute-force attacks impractical (see Sec. \ref{sssec:SeedGeneration}). In essence, this provides a maximum of security while keeping \nameit as user-friendly as possible.

As already mentioned, to set up a new device the user needs to transfer the seed \seed, the secret \ke, and the authentication token \tauth to the new device. All this data can be encoded together into a QR code \cite{qrcode}, a Base64 string, or transferred by manual file transfer. In particular the QR code-based approach is today well-known to users and very convenient.

\subsection{Summary}
\begin{figure}[t!]
	\centering
	\includegraphics[width=0.8\linewidth]{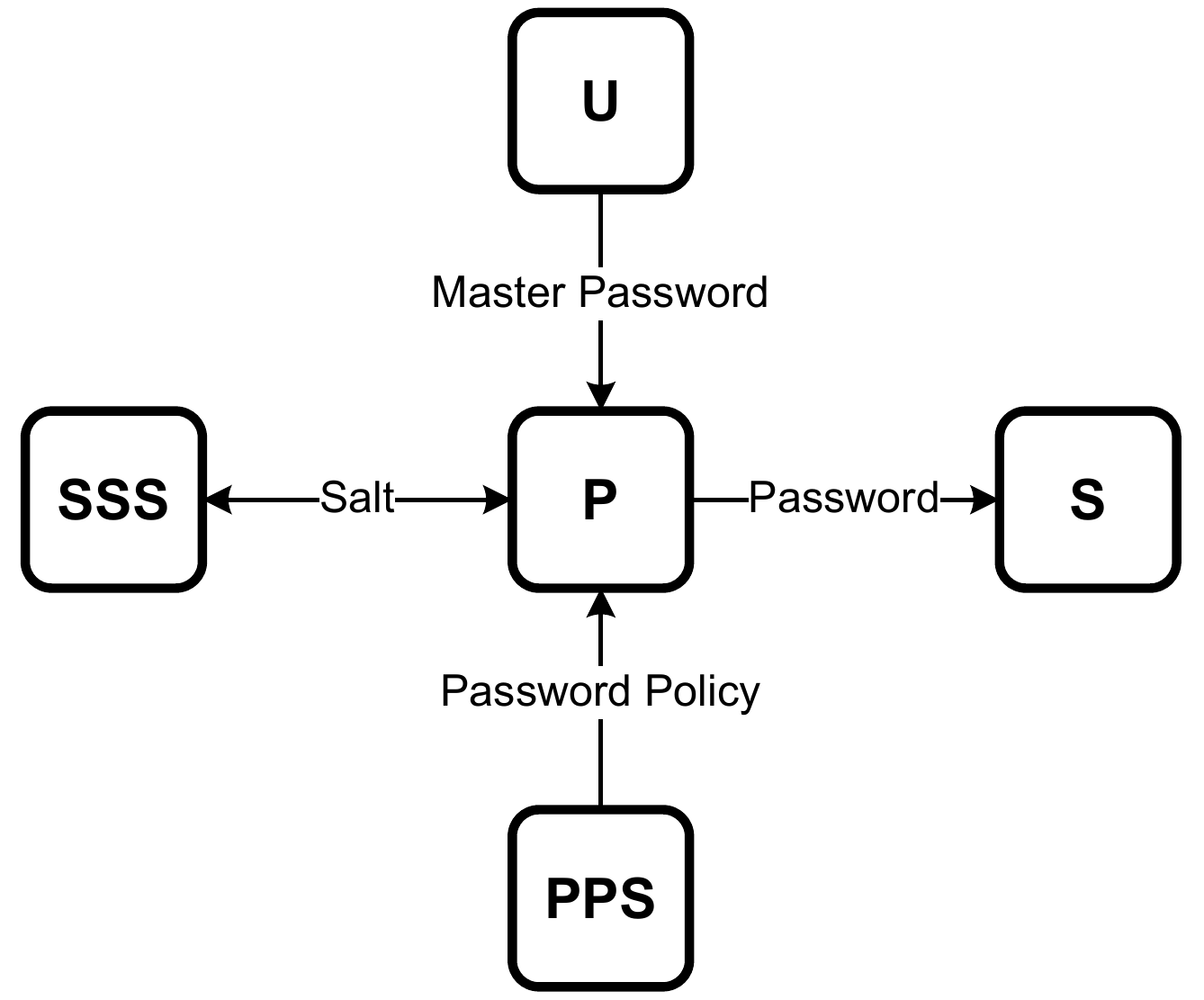}
	\caption{Overview. The user (U) wants to log in to the service (S). \nameit (P) prompts the user to enter his or her master password to access the seed. It retrieves the salt value from the Salt Synchronization Service (SSS) and the password policy from the Password Policy Service (PPS) for the service and finally computes the password.}
	\label{fig:infrastrucutre}
\end{figure}

In this section we described how \nameit works and introduced all involved participants. Figure \ref{fig:infrastrucutre} provides an overview and shows the information flow between the participants. The user (U) needs to enter his or her MPW to enable \nameit (P) to access the protected seed. To create a password for a service (S) \nameit determines the identifier of the service and requests the corresponding salt value and username from the Salt Synchronization Service (SSS) as well as the password policy from the Password Policy Service (PPS). Finally, \nameit computes the password.

\section{Implementation}
\label{sec:implementation}
The previous section described \nameit on a conceptual level. We now present more technical details regarding the implementation and used technologies.

We implemented \nameit as a proof-of-concept in Java and the \sss as well as the PPS as a Java web service. The \sss provides a SOAP-based interface for \nameit to create an account as well as to manage the account and the stored data. The implementation of the user authentication is based on Apache CXF and Bouncy Castle. The PPS provides a RESTful API for applications to retrieve the policies. The source code will be made available under an open-source license at www.palpas.info.

\subsection{Seed Generation and Protection}
\label{sssec:SeedGeneration}
The seed is generated gathering 256 bits of entropy from a strong random number generator (in our case the Java class \texttt{SecureRandom}). It is locally stored, encrypted with secret \kmpw using AES-256. The secret key \kmpw is derived from the user's MPW using the PBKDF2 function \cite{RFC2898}.

\subsection{Salt Generation and Synchronization}
\label{sssec:SaltValueSelection}
Using the URL of a service as identifier for the salt would raise privacy issues as argued above. Thus, \nameit generates identifiers as $id = \text{SHA-256}(\ke \| url)$ to protect a user's privacy. The parameter $url$ represents the URL of the service, in detail, the domain name, for instance \url{http://example.org}. 
The parameter \ke is a 256 bit key, generated during set-up and shared by all user devices. Prepending the secret \ke adds sufficient entropy to the input to make a brute-force search for a valid preimage, containing a URL intractable.

\subsection{Pseudorandom Generator}
\label{sec:ImplPseudorandomGenerator}
We implemented the PRG using AES in CBC mode. The seed \seed is used as key and the salt as input for the cipher. The ciphertexts form the pseudorandom output. We iteratively increment and then encrypt the salt value until enough random bits for the password generation are produced.

\subsection{Password Policies}
\label{sec:ImplPasswordPolicies}
We specified an XML-based data format for password policies to provide a standardized description of the password requirements. XML is widely supported by many platforms and programming languages. This ensures an easy adoption of our solution for other password tools. We provide a brief example of a password policy in Listing \ref{lst:policy}. The example policy allows passwords that have at least 6 and at most 12 characters. The policy specifies a character set of lower- and uppercase letters as well as digits and requires that the password must contain at least one digit. For instance, this policy allows passwords like \textit{Ha1K3A} and \textit{QSJe1Kf4qYt}.

\lstset{
	tabsize=2,
	label=lst:policy,
	caption={Password Policy. It specifies the minimum and maximum length of a password as well as the allowed character sets and additional restrictions.}
}
\begin{mylisting}
<PasswordPolicy>
	<MinLength>6</MinLength>
	<MaxLength>12</MaxLength>
	<CharacterSets>
		<CharacterSet name="LowercaseLetters">
			<Characters>abc...xyz</Characters>
		</CharacterSet>
		<CharacterSet name="UppercaseLetters">
			<Characters>ABC...XYZ</Characters>
		</CharacterSet>
		<CharacterSet name="Digits" minOccurrence="1">
			<Characters>0123456789</Characters>
		</CharacterSet>
	</CharacterSets>
</PasswordPolicy>
\end{mylisting}

\subsection{Password Generator}
\label{sec:ImplPasswordGeneration}
The PG maps the random value generated by the PRG to a password which complies with a given password policy (cf. Section \ref{sec:ConceptPasswordGeneration}). As example we consider a password policy that specifies a password length of 10 characters and a character set of uppercase and lowercase letters as well as digits. Thus, the character set has 26+26+10 = 62 characters in total. To derive a password of length $\ell$ using a character set of size $|character set| = \phi$, the PG takes $100+\lceil log_{2}(\phi^\ell)\rceil$ bits from the pseudorandom output and determines their decimal value modulo $\phi^\ell$. The resulting value is represented as $\ell$ base-$\phi$ numbers. In the above example the PG would take $100+\lceil log_{2}(62^{10})\rceil = 160$ bits, performs the modulo reduction, and splits the result in $10$ numbers between $0$ and $61$. Each base-$\phi$ number then serves as an index in the character set. For instance, $51=Z$ in our example character set. Using 100 bits more than required to get a number $>\phi^\ell$ guarantees an only negligibly biased distribution over the character set. More specifically, for each value $x$ in the character set, the probability that $x$ is chosen deviates from $1/\phi$ at most by $2^{-100}$. Please note that consequently adding fewer additional bits (e.g. 20 bits) recognizably increases the probability of small indexes to occur. Thus, it would produce biased passwords.

The password policy may specify additional restrictions, for example, that a password has to contain at least one digit. To support this setting while ensuring that passwords are uniformly distributed, we use rejection sampling. We generate a password form the pseudorandom output and check if it fulfills all restrictions. If not, the password is discarded and we use the next block of pseudorandom output from the PRG. The new password is generated and verified. This process is repeated until the password fulfills all restrictions and the verification succeeds. While this process in theory might never terminate, it terminates in practice after a few iterations. E.g., in our running example, the probability that in $\tau$ consecutive iterations a password is generated that contains no number is $((1 - 10/62)^{12})^\tau = 0.12^\tau < 2^{-3\tau}$. This probability vanishes exponentially fast in the number of iterations. Please note that a straight-forward approach of considering restrictions during the password generation (e.g.,\ generate $n-1$ characters, check if they contain a digit, if not select a digit for the $n$-th character) would produce biased passwords.

\subsection{Protecting the Usernames}
\label{sssec:username}
We use AES-256 in CBC mode with secret \ke to encrypt usernames before storing them on the \sss. Each username is encrypted separately using a different initialization vector (IV) to prevent that the same ciphertexts are produced for the same usernames. Both, ciphertext and IV are additionally protected using a Keyed-Hash Message Authentication Code (HMAC) \cite{DBLP:conf/eurocrypt/NamprempreRS14}. Ciphertext, IV, and HMAC are stored on the \sss for each service in addition to the salt value. The encrypted usernames can be retrieved by all a user's devices in the same way as the salts, using the identifier of the service.

Please note that for simplification we only consider that a user has one account at each service (i.e.,\ one username per service) for our prototype. However, our scheme can easily support several user accounts per service storing multiple salt values and encrypted usernames for the same identifier. \nameit retrieves this information as usually by requesting the data stored on the \sss for a particular identifier. Instead of returning a single salt value and username, the \sss would return a list of salt values with the corresponding encrypted usernames. \nameit decrypts the usernames and presents a selection dialog to the user.

\subsection{User Authentication}
\label{sec:ImplUserAuthentcation}
As already mentioned, the user authentication at the \sss is not password-based but uses public-key cryptography with a random authentication secret \kauth and corresponding public key. In detail, we are using TLS with client authentication for the user authentication at the \sss. Each user's device generates a new key pair comprising a public and a private key. During the account creation at the \sss, \nameit sends a Certificate Signing Request (CSR) \cite{RFC2986} to the \sss. The CSR contains only the public key and no further personal information. The \sss creates an X.509 certificate based on the CSR and transmits it to the device. The public key is bound to the account and stored in a list of trusted keys which are allowed to access and modify the salt values. Each user's device must add its own public key. To add an additional device an already registered device must request an authentication token \tauth from the \sss. The user must transfer the token to the new device. The new device creates its own key pair and a corresponding CSR that includes \tauth. Then it sends the token to the \sss. Hence, the authentication token ensures that only the user is able to add new devices.

Authenticating each device by its own public key has the advantage of fine-grained revocation. In case of theft or loss the user can revoke the access for a device by removing the public key from the list of trusted devices. Please see Section \ref{sec:SecurityDeviceTheft} for further details.

\section{Security Evaluation}
\label{sec:security}
In this section we provide a detailed security analysis of \nameit. We analyze the case of theft of a device, a phishing attack, security breaches at the services and the \sss, and finally a \sss failure.

\subsection{Resistance to Theft or Loss of User Device}
\label{sec:SecurityDeviceTheft}
Storing passwords or data to compute them, like \nameit does, bears the risk that devices are lost or get stolen, in particular in case of mobile devices. In that case, an adversaries might get access to the stored data. Although any mobile device is equipped with protection mechanisms such as passwords, PINs, or pattern locks, multiple surveys showed that only two third of the users use them \cite{DBLP:conf/soups/BruggenLKSCD13,DBLP:conf/ccs/EgelmanJPLCW14}. Even worse, unlock codes are often predictable and can be determined by simple guessing \cite{DBLP:conf/wisec/AndriotisTOY13,DBLP:conf/ccs/UellenbeckDWH13}. In essence, locking mechanisms for devices are an important aspect but one cannot rely on them as a protection mechanism for sensitive data.

As already described, \nameit uses a MPW to derive the secret \kmpw to encrypt the data stored on the device. However, this is only a first line of defense and increases the cost and effort for adversaries to obtain the data. It cannot be excluded that adversaries obtain the data carrying out a brute-force attack, especially if the user used a weak MPW. If encryption would be the only protection mechanism, users had to immediately change all their passwords because it would be just a question of time until adversaries find out the correct MPW. \nameit has a second line of defense.

\nameit computes a password using the seed and a salt. Either having the seed or the salt does not enable to compute the password. Thus, even if adversaries obtain the seed from a stolen device they need to get the salt values stored at the \sss to compute the user's passwords for the services. Each user's device uses a different authentication secret \kauth, in detail, a different secret and public key, for the authentication at the \sss. As described in Section \ref{sec:ImplUserAuthentcation} the \sss stores the public key of each device that is authorized to access the salt values. In case of a theft or loss, the user can revoke access for this device. Thus, if the MPW provides at least protection until the device's public key is revoked, the adversary is unable to retrieve the salts using the stolen device and consequently unable to compute the passwords. This gives the user enough time to change the seed on all devices and register the new passwords at all services.

\subsection{Resistance to Phishing Attacks }
\label{ssec:phishing}
Although password managers provide means against phishing attacks, they are also vulnerable to such attacks by themselves as soon as they use an online service, e.g., to synchronize passwords. The well-known procedure of luring users to fraud websites to gather usernames and passwords works also fine for password managers or synchronization services that use password-based authentication and provide a login through a web application. Our approach of using public key authentication or rather TLS client authentication is insusceptible against phishing attacks \cite{alsaid2006preventing,DBLP:conf/spw/Clayton05}, because there is simply no password that can be phished.

\begin{table*}[t]
	\begin{threeparttable}
		\renewcommand{\arraystretch}{1.2}
		\normalsize
		\small
		\centering
		\caption{Security Breaches}
		\label{table}
		\begin{tabular}{|p{.135\linewidth}|p{.04\linewidth}|p{.035\linewidth}|p{.115\linewidth}|p{.08\linewidth}|p{.08\linewidth}|p{.355\linewidth}|}\hline
			\bfseries Security Breach & \bfseries Seed & \bfseries Salt & \bfseries Salt Identifier & \bfseries Username & \bfseries Password & \bfseries Challenges to obtain the data \\ \hline\hline
			Service					& \no & \no  & \no $^{\ }$ & \yes $^{a}$ & \yes $^{a}$ & - Seed and salt must be brute-forced ($2^{512}$ bit) \newline - Various valid seeds and salts \newline - Password must be verified online \\ \hline
			SSS 					& \no & \yes & \yes $^{b}$ & \yes $^{b}$ & \no $^{\ }$ & - Seed must be brute-forced  ($2^{256}$) \newline - Secret \ke must be brute-force ($2^{256}$) \newline - Various valid seeds \newline - Password must be verified online  \\ \hline
			SSS and Service & \no & \yes & \yes $^{b}$ & \yes $^{b}$ & \yes $^{a}$ & - Seed must be brute-forced  ($2^{256}$) \newline - Secret \ke must be brute-force ($2^{256}$) \newline - Various valid seeds \newline - Password must be verified online  \\ \hline
		\end{tabular}
		\begin{tablenotes}
			\small
			\footnotesize
			\item Table 2 summarizes which (protected) data is obtained (depicted by $\bullet$) and not obtained (depicted by $\circ$) by adversaries in the respective security breach as well as which obstacles need to be overcome to break \nameit.
			\item $^{a}$ Maybe encrypted by the service (out of scope).
			\item $^{b}$ Encrypted with secret \ke.
		\end{tablenotes}
	\end{threeparttable}
\end{table*}

\subsection{Resistance to Security Breaches}
\label{ssec:securitybreach}
Numerous examples \cite{adobe,ebay,evernote,lastpasssn,twitter} tremendously showed the danger and impact of security breaches. Therefore, we analyze our approach regarding the threat of encountering a security breach at the \sss and/or at different services.

\subsubsection{Security Breach at the Service(s)}
In the first attack scenario we consider that adversaries get access to the password database of a service and in particular to the password of a user. The password was generated by the PG as described in Sections \ref{sec:ConceptPasswordGeneration} and \ref{sec:ImplPasswordGeneration}. The functionality of the PG and the policy are publicly available. Hence, adversaries are able to invert the PG and obtain (parts of) the pseudorandom output, generated by the PRG. However, it follows from the pseudorandomness of the PRG that knowing the output it is still computationally hard to learn (1) the seed and (2) the output of the PRG on input of the same seed but with a different salt value. Thus, adversaries are unable to compute other passwords from a stolen password.

The same applies in case that adversaries get access to password databases of multiple services and extract the passwords of a user at each service. As already mentioned, adversaries are able to obtain the corresponding PRG outputs, but they are unable to learn the seed or salt values.

Finally, we consider adversaries who are able to access the password database of a service multiple times, in particular after a password change. In detail, for an arbitrary time period the adversaries get a list of passwords generated by \nameit using the same seed but different salt values. Following the same argumentation as above, adversaries are able to invert the PG and obtain the corresponding pseudorandom output, but neither the seed nor the salts.

In essence, if adversaries are able to steal passwords, they cannot perform an offline attack to obtain information to compute other passwords. Due to the fact that our scheme generates an individual password for each service, adversaries can also not reuse the stolen password for other services. 

Please note that in theory adversaries can try all combinations of the seed and the salt and test if they generate the same password as stolen from the service. However, in practice there are a number of obstacles. First, the fact that the seed and the salt are 256 bit random values results in a search space of $2^{256}*2^{256} = 2^{512}$ bits. Second, even for a strong password policy that uses a 128 character set and length 20 passwords, there exist only $2^{140}$ possible passwords. Assuming that no further restrictions are applied and every seed-salt combination leads to a valid password, there are on average $2^{372}$ different seeds and salts that result in the same password. To find the correct seed-salt pair, an adversary has to verify that the seed and the salt are correct. The only way for the adversary to do this is to perform an online attack in which he computes a password for a different service and tests if it is accepted by that service. This is complicated by the fact that adversaries do not necessarily know which services a user uses. More importantly, services block accounts after a certain amount of wrong password attempts. Thus, even for computationally unbounded attackers an online attack is not applicable.

\subsubsection{Security Breach at the SSS}
In the second attack scenario we assume that adversaries are able to access the information stored at the \sss. In detail, the salt values, the encrypted salt identifiers, the encrypted usernames, and the public keys of the user's devices. The public keys for authentication are useless because the adversaries already have access to the \sss. The salts are randomly chosen values, independent of the passwords and do not reveal any information about the seed, the used services, or the corresponding passwords. To be precise, it follows from the pseudorandomness of PRG that only knowing the salt, guessing the output of PRG is infeasible. Therefore, adversaries are not able to generate any passwords. Thus, an offline attack is impossible.

Similar to the first scenario, in theory adversaries could brute-force the seed. This time, they even have the advantage of knowing a list of valid salt values. This reduces the search space to $2^{256}$ but it still forces adversaries to verify each seed in an online attack. In detail, for every seed and stolen salt the adversaries get a valid PRG output and password. Therefore, they need to verify the password with a service to check if they found the right seed. However, they do not know to which services the salts belong to and services restrict the number of login attempts. Using an additional brute-force attack against the secret \ke with complexity $2^{256}$ would allow an adversary to learn username and service URL. But they still need to test every seed online. Thus, an online brute-force attack is not feasible, even for computationally unbounded attacker. 

In summary, a security breach at the \sss might allow adversaries to brute-force the secret \ke which affects the user's privacy but does not reveal any information to obtain the passwords of the user.

\subsubsection{Security Breach at the Service and the SSS}
In the third attack scenario we assume a security breach at the service and the \sss. Adversaries would now obtain the password used at the service and the data stored at the \sss. As described in the first scenario, adversaries can compute the PRG output from the password by inverting the PG. However, it follows from the pseudorandomness of the PRG that it is still computationally hard to learn the seed in this scenario. 

An offline brute-force attack against the seed now becomes theoretically possible, because adversaries now have a password and a limited list of possible salt values. In case of a 256 bit seed as used by \nameit such a brute-force attack would take $2^{255}$ evaluations of the PRG. Moreover, such a brute-force attack would find many different valid seeds: Continuing the example from the first case, the attack would find $2^{116}$ seeds on average. Hence, again an online phase is required to verify every single seed from this set. This again leads to the same challenges as in scenario 2, i.e., either guessing the service for a salt or running a brute-force attack on the secret \ke. We noticed that knowing the service and the username would theoretical allow a known-plaintext attack. However, in practice, e.g.\ using ciphers like AES, that is not an issue and it leads back to a brute-force attack to obtain \ke. In summary, even in case of a security breach at the \sss and the service adversaries are unable to perform an offline brute-force attack.

\subsubsection{Summary}
The three attack scenarios are summarized in Table \ref{table}. It shows which data adversaries obtain in the case of a security breach in each attack scenario and what adversaries need to do to obtain the seed and break \nameit. The common approach of encrypting passwords allows adversaries to run offline attacks and to brute-force the encryption key. Furthermore, they can always verify if their attack was successful. Our approach does not allow this, because any possible seed (and salt) generates a valid password for a service. Thus, offline attacks are not possible. Adversaries are forced to choose a seed (and a salt), generate the corresponding password, and check if it is accepted by a service to know if they found the correct seed.

\subsection{Resistance to SSS failure}
The password generation is based on the seed and salt values stored on the \sss. Without being able to retrieve the salts form the \sss \nameit is not able to compute the password. Reasons for an unavailable \sss can be technical problems but also denial-of-service attacks. The reliability and fail-safe stability of the \sss can be ensured by redundancy of the hardware components to make technical problems very unlikely. Also denial-of-service attacks can be tackled by running multiple instances of the \sss on different locations, like common \textit{Content Delivery Networks} do. Furthermore, there exist a lot of research on detection and defense mechanisms of denial-of-service attacks \cite{DBLP:conf/eisic/MishraGJ11,DBLP:journals/comsur/ZargarJT13}. In summary, there exist multiple technical and organizational means to provide a reliable \sss.

It would have been possible to deploy a client-side solution to cope with the problem. The salt value needs only be changed if the password is changed and we can safely assume that this happens very rarely. Hence, \nameit could cache the salt values, which would allow generating the password even if the \sss is unavailable. However, this stands in direct contrast to the security advantage in case of theft or loss as described in Section \ref{sec:SecurityDeviceTheft}. Therefore, we decided against this solution and left it for future work.

\section{Conclusion and Future Work}
\label{sec:conclusion}
In this paper, we presented \nameit that generates strong and service-specific passwords and synchronizes them between the user's devices in a secure manner. We provided a detailed security analysis and showed among other things that (1) security breaches at the \sss and/or at the services do not allow adversaries to obtain the passwords, (2) the \sss is not vulnerable to phishing attacks because the user authentication is not based on passwords, and (3) even in case of a device theft the passwords are protected.  We also showed that the use of different secrets to provide data and privacy protection as well as user authentication improves the overall security and is still user-friendly. Users only need to remember a single master password. During setup of a new device some data has to be synchronized manually once. This can be done by scanning a QR code. We presented a solution that allows the password requirements of services to be automatically taken into account in the generation of strong passwords without any user interaction. The proposed standardized description, the automatic retrieval, and the distribution of password policies build a foundation and an open solution for various applications to include the password policies and to improve password-based authentication in the Internet. 

In the future, we plan to investigate a peer-to-peer approach for the salt synchronization in which the devices exchange the salt values directly. This has the benefit that we do not need a central server for the synchronization as done by the \sss in our case. Furthermore, we plan to analyze a machine-learning approach to generate the password policies automatically. We envision a tool that extracts the password requirements automatically from the services' websites and creates the corresponding password policies.

For future work one might also consider a trade-off between security, in case of theft or loss, and availability, in case of unavailability of the \sss, caching salts. Caching could be limited and the salts could be encrypted locally. \nameit could also be enhanced by more sophisticated authentication schemes like context-based authentication \cite{DBLP:conf/soups/HayashiDAHO13,DBLP:conf/eurossc/HulseboschBLEI07}. The user's context, e.g. the location, could be part of a sophisticated cache erasing algorithm which deletes the cache if the device leaves known locations. Furthermore, the caching could be enhanced by pushing the salt values instead of fetching them on demand. More precise, if a user's device updates the salt at the \sss, the \sss could automatically push the new value to the other devices. This way all devices have the latest salts even if the \sss was unavailable with high probability. A sophisticated caching mechanism could protect the user in case of device theft or loss as well as guaranteeing availability even if the \sss is unavailable.

\section{Acknowledgments}
This work was supported in part by the European Union within the 7FP project FutureID (ICT-318424) and by the Netherlands Organisation for Scientific Research (NWO) under grant 639.073.005.

\bibliographystyle{abbrv}
\bibliography{references,nist,rfc}

\appendix
\section{Protocol Flows}
This appendix provides detailed figures and descriptions of the protocol flows for the setup of \nameit, registration of a new user device, synchronization of an additional password, login to a service, and finally the update of a password. The following diagrams contain the user (U), the Salt Synchronization Service (SSS), \nameit (P) running on a user's device, and an exemplary service (S).

\subsection{Setting up PALPAS}
\begin{figure}[t]
	\centering
	\includegraphics[width=\scale\linewidth]{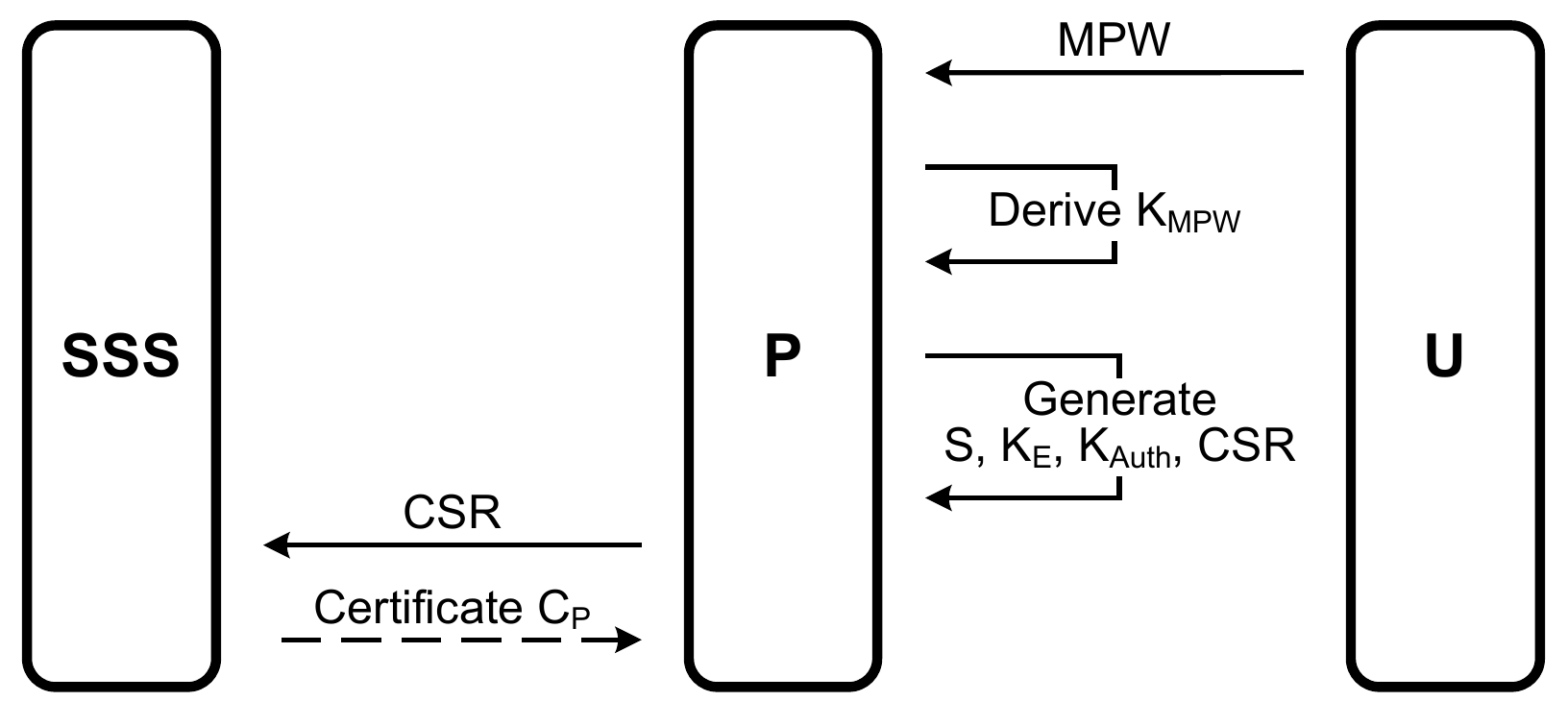}
	\caption{Setting up PALPAS.}
	\label{fig:flow_setup}
\end{figure}

After installing and starting \nameit on a device the user needs to select and enter a master password (MPW) (cf. Figure \ref{fig:flow_setup}). \nameit derives the secret \kmpw from the MPW and uses it to encrypt any data that will be stored on the device (e.g. the seed). Subsequently, \nameit generates the seed \seed, the secret \ke, the secret \kauth, and a corresponding Certificate Signing Request (CSR). It sends the CSR to the \sss in order to create a new account for the user. The \sss creates a certificate based on the CSR and sends it to PALPAS. The certificate is used to authenticate the user to the \sss. More precise, PALPAS uses the certificate for a TLS client authentication to the \sss.

\begin{figure}[b]
	\centering
	\includegraphics[width=\scale\linewidth]{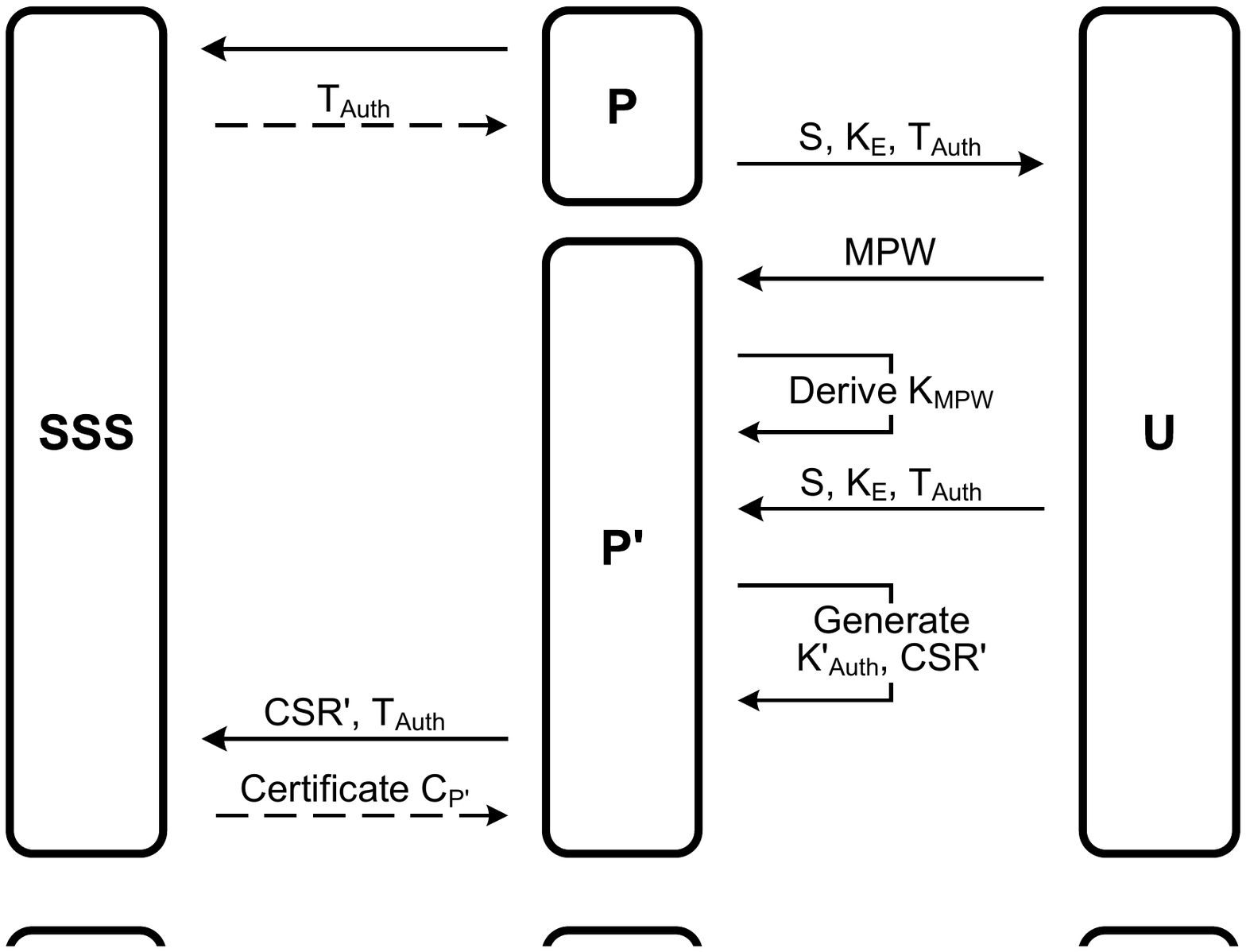}
	\caption{Adding a new user device.}
	\label{fig:flow_addingdevice}
\end{figure}

\subsection{Adding a Device}
The procedure to use an additional device with \nameit comprises two steps: First, the user requests an authentication token from the \sss using an already registered device. Second, the user registers his or her additional device at the \sss using the authentication token. As depicted in Figure \ref{fig:flow_addingdevice}), \nameit (P) requests a token \tauth from \sss and encodes it together with the seed \seed and the secret \ke as a QR code (or stores it in a file). The user installs \nameit (P') on the new device and enters a master password. Please note that the master password is only used to protect the local data. Therefore, the user can choose different master passwords for the devices. Subsequently, the user transmits the seed \seed, the secret \ke, and the token \tauth to the new device (e.g. scanning the QR code or by manual file transfer). \nameit (P'), running on the new device, generates a key pair and a corresponding CSR'. It sends the CSR' and the token \tauth to the \sss to register the device. Finally, the \sss creates a new certificate for P' and sends the certificate to it.

\subsection{Synchronizing a new Password}
\begin{figure}[h]
	\centering
	\includegraphics[width=\scale\linewidth]{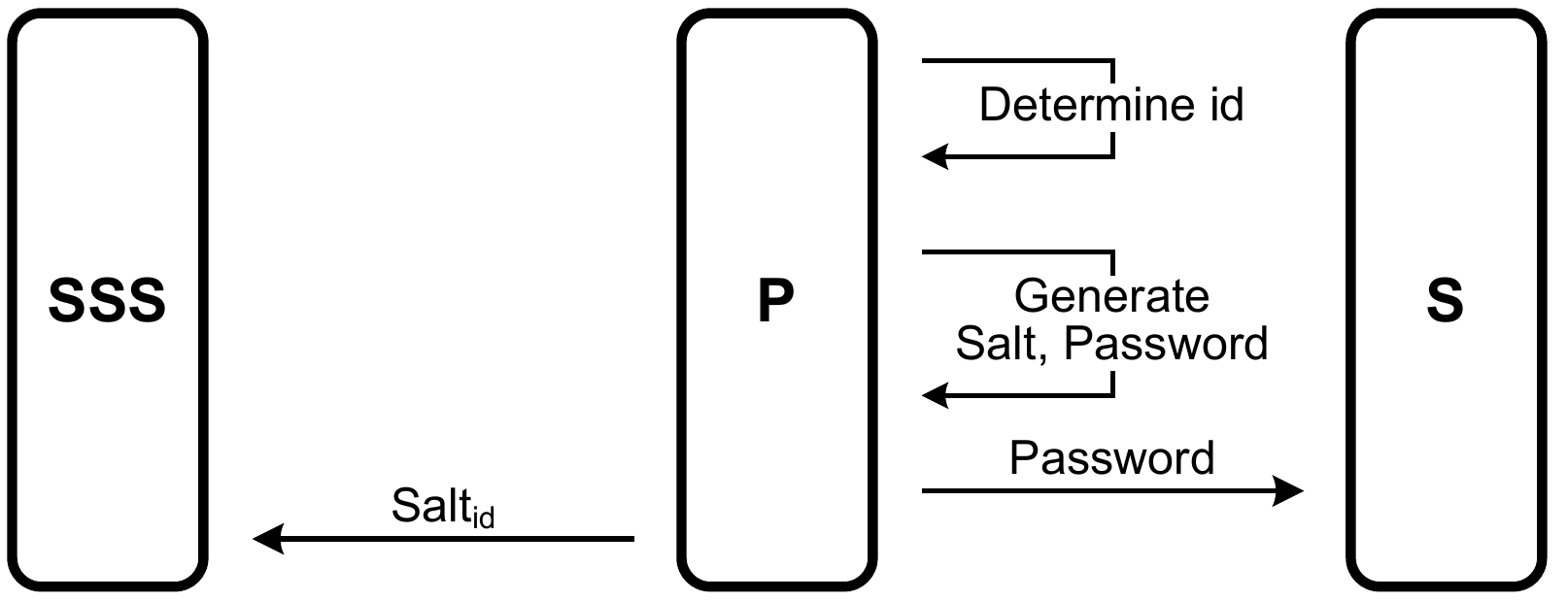}
	\caption{Synchronizing a new password.}
	\label{fig:flow_addingservice}
\end{figure}

Figure \ref{fig:flow_addingservice} shows the synchronization of an additional password with \nameit. For instance, in case the user wants to use a new service. As a first step, the identifier (id) of the service is computed (cf. Section \ref{sssec:SaltValueSelection}). Subsequently, \nameit generates a salt and the corresponding password for the service. The user sets the password at the service and \nameit stores the salt at the \sss associated with the identifier and the encrypted user data. Other user's devices are able to generate the same password after synchronization with the \sss.

\subsection{Logging into a Service}
\begin{figure}[h]
	\centering
	\includegraphics[width=\scale\linewidth]{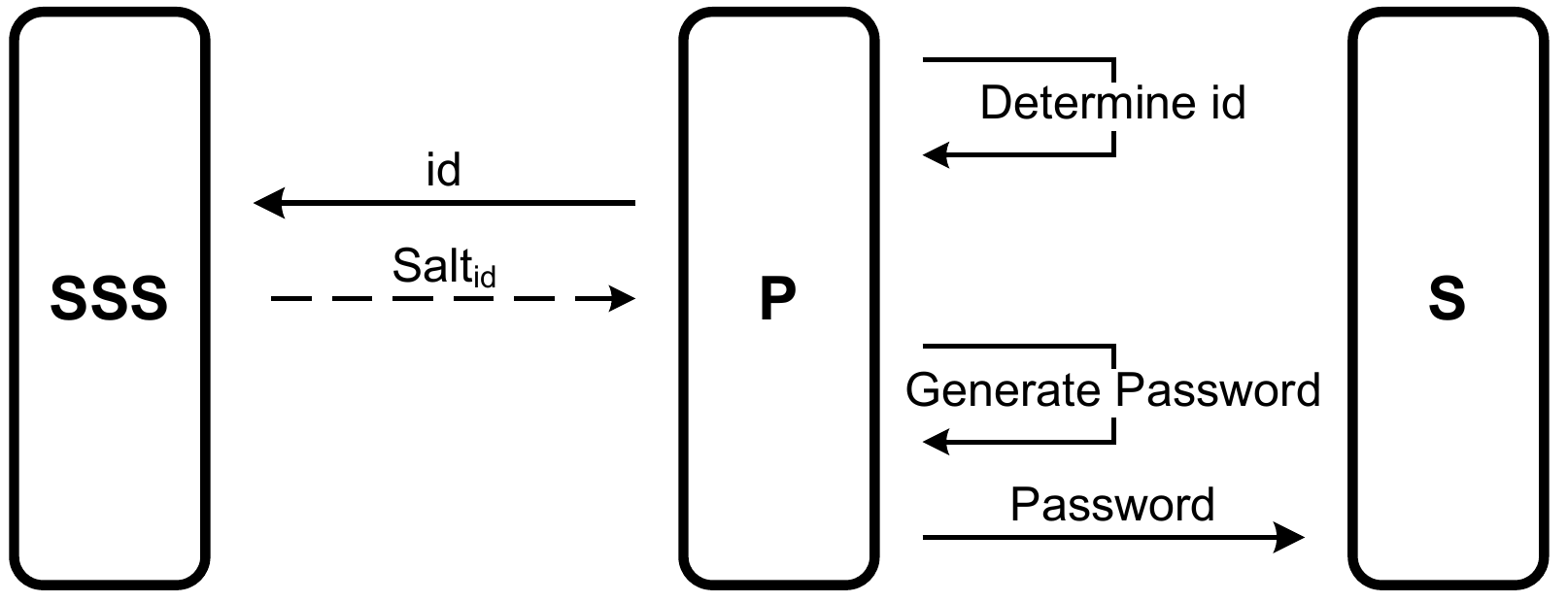}
	\caption{Logging into a Service.}
	\label{fig:flow_login}
\end{figure}

If the user wants to login to a service, \nameit determines the identifier of the service (cf. Figure \ref{fig:flow_login}). Then \nameit requests the corresponding salt from the \sss and computes the password for the service. Finally, the login is performed and the user has access to the service.

\subsection{Updating a Password}
\begin{figure}[h]
	\centering
	\includegraphics[width=\scale\linewidth]{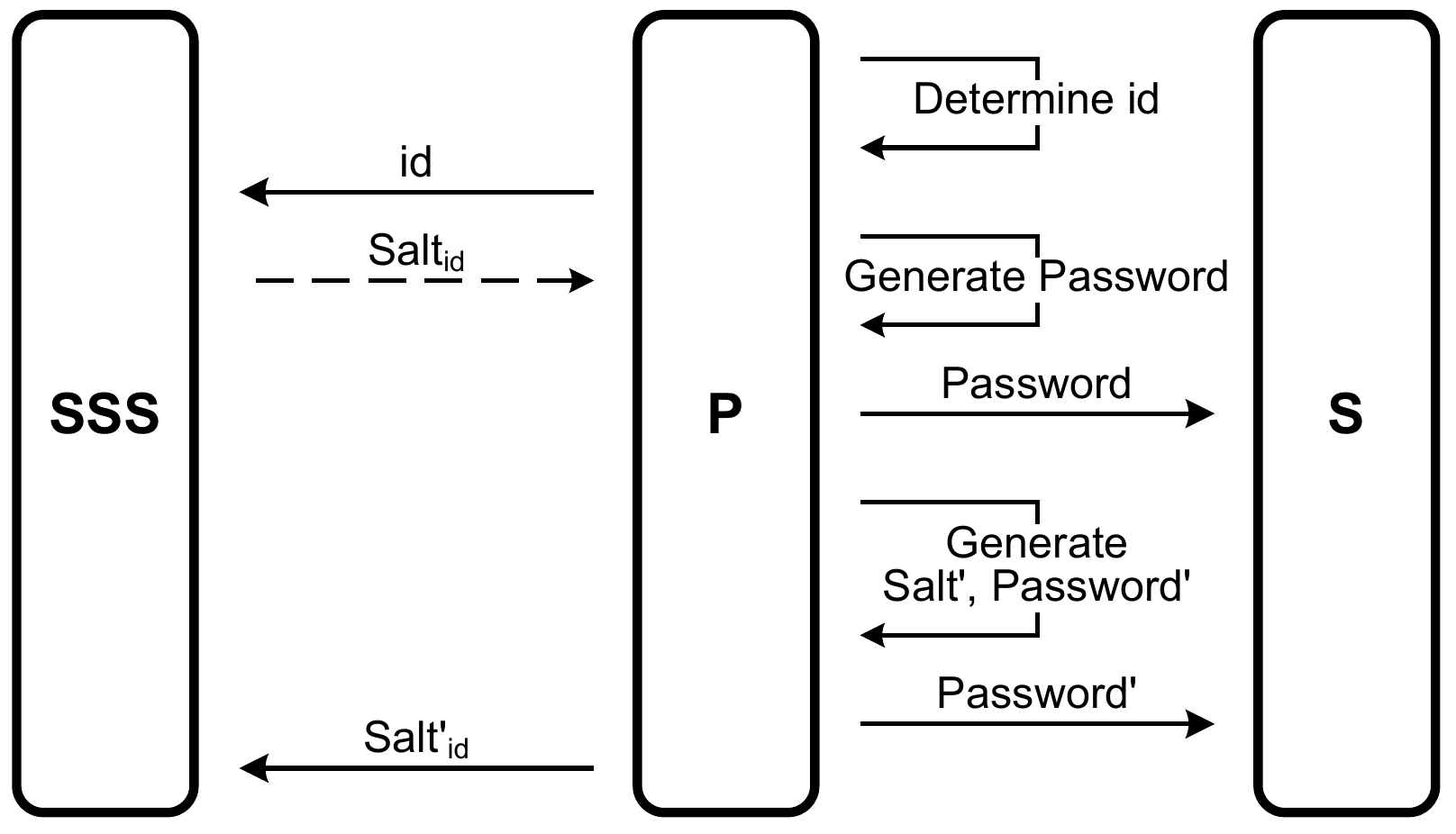}
	\caption{Updating a Password.}
	\label{fig:flow_pwchange}
\end{figure}

Updating a password which is synchronized by \nameit is a combination of the procedures performed for the login and the synchronization of a new password. As illustrated in Figure \ref{fig:flow_pwchange}, first, \nameit determines the identifier of the service. Second, it retrieves the corresponding salt by the \sss. Third, it generates the password and the user logs into the service. Fourth, \nameit generates a new salt and the corresponding password. Fifth, the user updates the password at the service. Finally, if the password change at the service was successful, \nameit stores the new salt value at the \sss using the same identifier as before.

\end{document}